
\input harvmac
\def \BW {\overline W }
\def \sn {\sqrt N }
\def \ra {\rightarrow}
\def \inf {\infty }
\def \bt {\overline \tau}
\def \bl { \overline \l}
\def \tmu {\tilde \mu }
\def \dvps {{\dot \varphi}^2}
\def \dls {{\dot \lambda }^2}

\def \bvp {\overline \vp}
\def \dlis {{{\dot \l}_i}^2}

\def \dop  {\dot {\phi} }
\def \ddp {\ddot \phi }
\def \L {\Lambda}
\def \dL {\dot \Lambda}
\def \ddL {\ddot \Lambda}
\def \dr {\dot \rho}

\def \sps {\psi^* }
\def \stps {{\tilde \psi}^* }

\def \eq#1 {\eqno{(#1)}}
\def \e {\rm e}

\def \tm {\tilde m }
\def \summ {\sum_{m=1}^{\infty}}
\def \sumtm {\sum_{\tm =1}^{\infty}}

\def \evp {\rm e^\varphi }

\def \ra {\rightarrow }
\def \e#1 {{\rm e}^{#1}}
\def \ps {\psi}
\def \tps {\tilde \psi}
\def \tp {\tilde \phi}
\def \tvp {\tilde \varphi}
\def \tl {\tilde \lambda}
\def \mi {m_{i}^2}
\def \tm  {\tilde m}
\def \tmi {{\tilde m}_{i}^2}
\def \dps {\dot \psi}
\def \ddps {\ddot \psi}
\def \dtps {\dot {\tilde \psi}}
\def \ddtps {{\ddot {\tilde \psi}}}
\def \dpss {{\vert {\dot \psi} \vert }^2}
\def \pss {{\left| {\psi} \right|}^2}
\def \tpss {{\vert {\tilde \psi} \vert}^2}
\def \dtpss {{\vert {\dot{\tilde \psi}} \vert }^2}

\def \a {\alpha}
\def \b {\beta}
\def \sh {\ {\rm sinh} \ }

\def \th {\ {\rm tanh} \ }
\def \ln {\rm ln }

\def \l {\lambda}
\def \p {\phi}
\def \vp {\varphi}
\def  \g {\gamma}
\def \o {\omega}
\def \r {\rho}
\def \k {\xi}

\def \s {\sigma}

\def \b {\beta}

\def \pa {\partial}

\def \sqG {\sqrt {-G}}
\def \Goo {G_{00}}
\def \Gmn {G_{\mu \nu}}

\def \CE {\cal E}
\def \CU {\cal U}

\def \dl {\dot \lambda}
\def \ddl {\ddot \lambda}
\def \dvp {\dot \varphi}
\def \ddvp {\ddot \varphi}
\def \sm {\sum_{i=1}^{N}}

\def\NP {  Nucl. Phys.\ }
\def \PL { Phys. Lett.\ }
\def \MPL { Mod. Phys. Lett.\ }
\def \PRL {  Phys. Rev. Lett.\ }
\def \PR  { Phys. Rev. \ }

\Title{DAMTP - 37 - 1991}
{\vbox{\centerline{Dilaton, winding modes and }
\vskip2pt\centerline{ cosmological solutions}}}

\centerline{ A.A. Tseytlin
\footnote{$^\dagger$}
{On leave of absence from the Department of Theoretical Physics, P. N.
Lebedev Physics Institute, Moscow 117924, USSR.
e-mail: aat11@amtp.cam.ac.uk }}
\bigskip
\centerline{\it DAMTP}
\centerline{\it Cambridge University}
\centerline{\it  Cambridge CB3 9EW , U.K.}
\baselineskip=20pt plus 2pt minus 2pt
\vskip .3in


We review some formal aspects of cosmological solutions  in  closed string
 theory
 with duality symmetric ``matter'' emphasizing the necessity to account for
  the dilaton dynamics for a proper incorporation of duality.
 We consider two models : when the matter action is
  the classical action of the fields
corresponding to  momentum and winding modes and when the matter
action is represented by the quantum  vacuum energy of the string
compactified on a torus. Assuming that effective vacuum energy
is positive one finds that in both cases  the scale
factor undergoes  oscillations from maximal to minimal
values with the amplitude of oscillations decreasing to zero or
increasing to infinity depending on whether the effective coupling
(dilaton field) decreases or increases with time. The contribution of
the winding modes to the classical action
prevents infinite expansion. Duality is ``spontaneously
broken'' on a solution with generic initial conditions.

\bigskip
\it{Submitted to Classical and Quantum Gravity }
\Date{10/91} 


\newsec{ Introduction}

Target space duality is known to be one of fundamental properties
of  string theory [1]. Being in some sense a symmetry between large
and small scales it may play an important role in string cosmology
[2] if the contributions of the winding modes are included
[2-4].  In order for the string theory cosmological equations to be
invariant under the ``non-static'' generalisation of duality
(or `` $\s$ model duality'') [5]
 it is necessary to account for the dependence on the dilaton field
which transforms under the duality [5-10]. It is the combined
metric-dilaton system of equations that is invariant under the
duality if the ``matter'' action is duality symmetric (i.e.
contains contributions of both momentum and winding modes) [11].

In this paper we shall discuss some formal aspects of  ``cosmological''
(i.e. time dependent) solutions  in the case of duality
symmetric ``matter'' with one of the aims being to illustrate
 the effect of the
winding mode contributions on the evolution of the scale factor.
We shall not  consider ``realistic'' cosmological
scenarios  as this was already done in [12]. The present paper is a review and
an extension
of certain aspects discussed in [12].

Our choice of the matter actions will be based on the assumption of
``adiabaticity'' understood in the following sense. Let the
space-time be a product of a time line and an $  N $ dimensional torus
with  radii $a_i=\rm e^{\l_i} $ (which we shall usually take to be
equal). We shall assume that the matter action can be represented
by the classical action of the modes of the string theory
compactified on the torus with $a_i$  replaced by functions
of time. The resulting action will be invariant under the duality
transformation which inverts the radii, shifts the dilaton and
interchanges momentum and winding modes. The corresponding
equations will describe evolution of time dependent perturbations
of the torus vacuum (from a different point of view these
equations may be considered as a  generalisation of the Kosterlitz--Thouless
type renormalisation group equations [13]).

We shall also consider the case when the matter action is
represented by the (zero temperature) vacuum energy of the gas
of momentum and winding modes (i.e. by the
partition function of the string compactified on the
torus) which is automatically invariant under the duality
transformation of $a_i$ and the dilaton. Though the vacuum energy vanishes
for the superstring  a formal analysis  of this case  is useful
since the resulting equations are similar to the system one finds in a
more realistic finite temperature case [12].

We shall try not to
specify the number $ N$ of space dimensions (i.e. keep the value of
the central charge $c\sim D-26 (=N-25 ) $ arbitrary ) so that some of our
results may apply, e.g.,  to the case of  $D=2$  string theory (
 cosmological solutions with non-zero $c$  in the matter-free case
were discussed in [14-15]).

We shall find that both in the ``classical'' and ``quantum'' matter
cases the scale factor (the common radius of the torus) oscillates between
finite maximal and minimal values .  If  the effective dilaton coupling is
decreasing with time the amplitude of oscillations  is also
decreasing and  the asymptotic value of the radius is the
Planck one ($\sqrt {\a'}$). Though the duality is spontaneously broken by
initial conditions it is thus restored asymptotically. The role of
the winding mode contributions is to prevent an infinite expansion in
the ``classical'' matter case  [12] (and,
modulo the issue of interpretation, to prevent a contraction to zero in
the ``quantum'' matter case).

We shall describe the expansion/contraction in terms of  the
original unrescaled (``$\s$ model'') metric. Though the effective
gravitational constant is then time dependent, this is the
metric  strings directly interact with and hence the one measured by
stringy ``rods''. Similar point of view seems to be expressed
in refs. 16--17 where the question of  which metric
should be used to measure the expansion in string cosmology was discussed.
 The use of ``$\s $ model '' metric
is particularly natural in the duality symmetric case
since it is the unrescaled metric that has a simple transformation law
under the duality. We shall assume that the dilaton does not have a
 non-perturbative
potential, i.e. is ``massless''.

In Sect.2 we shall present the basic ``cosmological'' equations for
the two choices of the matter sources. In the case of the classical
matter we shall consider a ``mechanical'' interpretation of the
resulting  system of equations (with the central charge playing the
role of an ``energy'') and also note a correspondence with the
renormalisation group equations (for a related discussion of the RG
equations see [18]).

In Sects.3 and 4 we shall study the behaviour of the solutions
 in the classical and quantum matter cases
respectively. The qualitative analysis shows that the
radius is oscillating with time between its extremal values with the
amplitude of oscillations increasing to infinity or decreasing to zero
 depending on initial conditions for the dilaton.

Sect.5 contains some concluding remarks.
We shall briefly discuss the ``anisotropic'' case
in which there are ``large''
 and ``small'' (internal) radii .

In Appendices A and B  we shall  explicitly solve the basic ``classical'' and
``quantum'' systems of equations in the asymptotic region of large
radius and demonstrate the existence of a maximal radius of expansion.

\newsec{Basic equations}

1.  We shall consider only the leading order terms in the low energy
expansion of the tree level gravitational effective action of the
closed bosonic string theory [19]
(we shall absorb the gravitational coupling constant into $\p$ and
often set $\a'$=1)
$$ S_0 = - \int d^D x \sqG \  \e{-2\p}   \ [ \ c + \ R \ + 4 (\pa \p )^2
\ ] \ , \ \eq{2.1} $$
$$ c= - {2\over 3\a'} (D-26)   \  . $$
The corresponding equations for the gravity plus matter action
$S=S_0 +S_m \ $    are
$$ R_{\mu \nu} + 2 D_{\mu} D_{\nu} \p  - \half  G_{\mu \nu} \CL
= { \e{2\p}  \over
\sqG  } {\delta S_m \over \delta G^{\mu \nu}} \ \ , \eq{2.2}
 $$
$$ \CL \equiv c+ R + 4 D^2 \p - 4 (\pa \p)^2 = - \half
{\e{2\p} \over \sqG} {\delta S_m \over \delta \p} \ \ . \eq{2.3} $$
We shall consider the following cosmological background
$$ ds^2 = - dt^2 + \sm a_i^2 (t) dx_i^2  \  \  , \eq{2.4} $$
$$ a_i= {\rm e}^{\l_i(t)} \ , \ \ \p=\p(t) \ , \ \ N=D-1\ \ . $$
For a proper interpretation of the matter action we shall discuss
it is necessary to assume that at least some of coordinates $x_i$
are periodic.

It is useful to introduce the ``shifted'' (and rescaled by 2) dilaton
field  $\vp$
$$ \vp \equiv 2\p - \sm \l_i \ \ ,\  \ \ \sqG\   \e{-2\p} = \e{-\vp} \   \ ,
\eq{2.5} $$
which is invariant under the duality transformation [6-10]
$$ \tl_i=-\l_i \ \ , \ \tp =\p - \l_i \ \ , \ \  \tvp=\vp \eq{2.6} $$
(here $i=1,...,N$ ; more general duality transformations are
obtained by combining these basic ones ).
 We are assuming that all the dimensions are compact.
In general, the effective coupling is represented
by the   dilaton  (2.5)  where only the logarithm of the volume of
the compact part of the space is subtracted out.
 Re-written in terms of
$\l_i$ and $\vp$ the action and equations (2.3),(2.2) take the form
(we drop a constant factor of integral over $x_i$)
$$ S=-\int dt\  \e{-\vp} \sqrt {-\Goo}\  [ \ c - G^{00} \sm \dl_i^2
 + G^{00}\dvp^2 \ ]  + S_m [G_{00}\ , \l_i \ , \vp] \ \ , \eq{2.7} $$
$$S_0= - \int dt \ \e{-\vp} [\  c + \sm \dl_i^2 - \dvp^2 ]\ \ ,
\eq{2.8} $$
$$ c-\sm \dl_i^2 +\dvp^2 = -2 \e{\vp} {\delta S_m \over \delta G_{00}} \ ,
\eq{2.9} $$
$$\ddl_i - \dvp \dl_i = -\half \e{\vp} {\delta S_m \over \delta \l_i} \ ,
\eq{2.10} $$
$$ \ddvp - \sm \dl_i^2 = - \e{\vp} ( {\delta S_m \over \delta G_{00} } -
\half
{\delta S_m \over \delta  \vp } )  \ .\eq{2.11} $$
Similar set of equations was discussed in [10-12].
Eqs.(2.9)-(2.11) follow from the action (2.7) (after the variation
$G_{00}$ is set equal to --1). If one stars from the action (2.8)
(where $G_{00}$ =--1) one
should add  eq.(2.9) as an additional ``zero energy'' constraint.
 If the original matter action is generally
covariant the time derivative of (2.9) vanishes identically once
(2.10) and (2.11) are satisfied. The examples of matter
actions $S_m[G_{00},\l_i,\vp]$ we shall consider  will be invariant
under reparametrisations of time so
that the identity
relating the derivative of (2.9) to (2.10),(2.11) will be satisfied.
Hence one of eqs.(2.10),(2.11) is redundant and can be dropped
(if  $\dl \not= 0\ , \dvp \not= 0$). The meaning of eq.(2.9) is that
the integration constant  which appears after integrating once
eqs.(2.10),(2.11) is not arbitrary but is proportional to $c$.

Eqs.(2.7)--(2.11) are invariant under the duality transformation
(2.6) provided the matter action is duality symmetric. Given a
solution of (2.9)--(2.11) we can then generate other solutions by
applying duality transformations.

Let us note that in the duality non-invariant case  when the matter
action is given by the dilaton potential $ V(\p) $ , i.e. (we shall assume
the isotropic case when  all radii are equal)
$$ S_m = \int \ dt \ \sqrt  {-\Goo} \ V(\vp + N \l ) \ \e{ N\l} $$
the system of equations (2.9)--(2.11) reduces to
$$ c- N\dls + \dvps = \e{\vp} V  \ \ , $$
$$  \ddl - \dvp \dl = - \ha \e{\vp} ( V + V' ) \ \ , $$
$$  \ddvp - N \dls = \ha \e{\vp} (V+ V')  \ \ . $$

2. As was noted in the Introduction, one of our aims is to understand the
effect of the winding modes on cosmological evolution. Let us first
consider the case when  the matter action is represented by the
classical action of the fields corresponding to momentum and
winding modes of string theory compactified on a torus. For
simplicity we shall consider just two complex scalar fields
representing the momentum and winding modes of the tachyon field
($\tilde x_i$ are the ``dual'' coordinates ; $\a'=1$)
$$T(x,\tilde x)= \ps\  \exp {(i\sm m_i x_i)} +\tps \ \exp{(i\sm \tm_i \tilde
x_i)} +  c.c + ...\ , \ \eq{2.12} $$
$$ S_m = \half \int dt\  \e{\vp} \sqrt {-G_{00}}\  [\  G^{00} \dpss +
(\sm \mi \e{-2\l_i} -4 ) \pss $$
$$ + \ G^{00} \dtpss + ( \sm \tmi \e
{2\l_i} -4 ) \tpss  + 2 U (\ps,\tps)\ ] \ , \ \eq{2.13} $$
where the potential $U$ starts with quartic terms (cubic interactions
involve other modes  as well).  To desribe the case when
$\ps$ and $\tps$ represent higher level scalar
modes the   tachyonic mass term --4 should be
replaced by   $m_0^2= 4(n-1) \  ,\  n=0,1,...$ .
Variation of the total action (2.7), (2.13) over $G_{00} ,\  \l_i,\  \vp,\
\ps,\ \tps$ gives the following system of equations (cf.(2.9)--(2.11)
; $G_{00}$ = --1)
$$ c- \sm \dl_i^2 + \dvp^2 = \half \  [\  \dpss + ( \sm \mi \e {-2\l_i}
-4)\pss $$
$$+ \dtpss + (\sm \tmi \e {2\l_i} -4)\tpss + 2U \ ] \ ,
\eq{2.14} $$
$$ \ddl_i - \dvp\dl_i = \half ( \mi \e{-2\l_i} \pss - \tmi \e{2\l_i}
\tpss ) \  , \eq{2.15} $$
$$ \ddvp - \sm \dl_i^2 = \half ( \dpss + \dtpss ) \ \ , \eq{2.16} $$
$$ \ddps - \dvp  \dps + ( \sm \mi \e{-2\l_i}  - 4 ) \ps
+ {\pa U \over \pa  \sps} = 0  \ ,
\eq{2.17} $$
$$ \ddtps - \dvp \dtps + ( \sm \tmi \e{2\l_i} - 4 ) \tps +
{\pa U \over \pa \stps } =0 \ .
\eq{2.18} $$
Eqs.(2.14)--(2.18) are invariant under the duality transformation
which is the combination of the transformation (2.6) of ``moduli''
and of the interchanging of the momentum and winding modes,
$$ \l_i \rightarrow -\l_i \ , \ \ \vp \rightarrow \vp \ , \ \
\ps \leftrightarrow \tps \ , \ \ m_i \leftrightarrow  \tilde m_i
\ . \eq{2.19} $$
The system (2.14)--(2.18) is consistent in the sense that the time
derivative of the ``zero energy'' constraint (2.14) is proportional
to eqs.(2.15)--(2.18). In fact, eqs.(2.14)--(2.18) have the following
``mechanical'' interpretation. Eqs.(2.15), (2.16) (combined with
(2.14))  and  (2.17), (2.18) follow from the action (2.7), (2.13)
(with $G_{00}=-1$)
$$S=  - \int dt (\  K \ - \ \CU\ ) \ \  , \eq{2.20} $$
where the kinetic and potential terms are
$$ K = \e{-\vp} ( \ - \dvp^2 + \sm \dl_i^2  + \half \dpss + \half
\dtpss \ )\ \ ,  $$
$$ {\CU } ={ \e{-\vp} } \  [ \  - c + \half
 ( \sm \mi \e{-2\l_i} -4 ) \pss + \half (
\sm \tmi \e {2\l_i} - 4 ) \tpss +  U  \ ] \ . \eq{2.21} $$
The action (2.19)--(2.21) describes a mechanical system with the
$M=1+N+4$  dimensional  configuration space with the Minkowski
signature metric $g_{AB}= \e{\vp} \eta_{AB} \ , \eta_{AB} =
(-,+,..., +)$  and the dilaton $\vp$ playing the role of a time
coordinate . This space is curved. For example, its scalar
curvature  is $R= {1\over 4} (M-1)(M-2) \e{-\vp} $ ; it vanishes
 if $N=1$ ( i.e. $D=2$)  and the matter fields are absent.
Eq. (2.14) is then recognised as the zero energy condition   $K+\CU$ =0
which should be added ``by hands''.

It is interesting to note that eqs.(2.15), (2.17), (2.18) generalise
the renormalisation group equations for the couplings which
correspond to perturbations of the torus vacuum by marginal
operators. Let us consider for simplicity the case of one
dimensional torus ($N=1$). The action which generates the
corresponding $\b$ - functions is given by  [20]
$$ L = \int dx d {\tilde x} \  [\ \half G^{-1} ({\del T \over \del x})^2
+ \half {\tilde
G}^{-1} ( {\del T \over \del \tilde x } )^2 - 2T^2 + {\textstyle {1\over 6}
}T^3 +... ] \ , \
\eq{2.22} $$
$$ \tilde G=G^{-1}  \ ,\ \ \ G=\e{2\l} \ , \ $$
$$ \b^{T}= \dot T = \half {\delta L \over \delta T }\
, \  \ \b^G= \dot G = - \half
{\del L \over \del G^{-1} } \ . \eq{2.23} $$
The rotational invariance constraint
${{\partial}^2 T\over
\partial x \partial \tilde x }=0 $
implies
$$ T= \psi (x) + {\tilde \psi
} ( \tilde x   ) = {\sum_{m=-\infty}^{\infty}} \psi_m \ \e{imx}
+ {\sum_{\tm =- \infty}^{\infty}} {\tilde \psi}_{\tilde m} \ \e{i\tm \tilde x
} \
, \eqno{(2.24)} $$
so that the action and RG equations take the form [13,20]
$$ L = \half \summ ( G^{-1}m^2 -4) \vert {\psi}_m \vert^2 + \half \sumtm
(G \tm^2 -4 )\vert {\tilde \psi}_{\tm} \vert^2 + O(\psi^3 , {\tilde
\psi}^3 ) \ , \eq{2.25} $$
$$ {\dot \psi}_m= \ha{\del L \over \del
\psi_{-m}} = {1 \over 4 } (G^{-1} m^2 -4 ) \psi_m +  O(\psi^2)
\ , \eq{2.26} $$
$$ {\dtps}_{\tm}= \ha{\del  L \over {\del
{\psi_{-\tm}}}} = {1 \over 4 } ( G\tm^2 -4 ) {\tilde \psi}_{\tm}
+ O ({\tilde \psi}^2) \ , \eq{2.27} $$
$$ \dot G= - \ha { \del L \over \del G^{-1}} = - {1 \over 4 } ( \summ m^2
 \vert \psi_m \vert^2
- G^2 \sumtm {\tm}^2 \vert {\tilde \psi}_{\tilde m} \vert^2  ) \ .  \eq{2.28}
$$
Here $\psi_m \ ,\  \tilde \psi_{\tilde m } \ ,\ \ G= \e{2\l} \ $ are
``running''
 couplings ,
and the derivatives are taken with respect to the RG parameter
$\tau$. The correspondence between eqs.(2.26)--(2.28) and
(2.14)--(2.18) is established as follows. If we drop the second
derivative terms in (2.15), (2.17), (2.18), set $ N=1$ ,
 $ \dvp $=4  and keep just two modes ( $m$ and $\tm$) then eq.
(2.15) becomes identical to  (2.28)  while eqs. (2.17), (2.18)
reduce to (2.26), (2.27). As discussed in [18] this prescription corresponds
to taking a  semiclassical limit  if the cosmological equations
(2.14)--(2.18) are
interpreted as RG equations in the presence of quantized 2d
gravity. To satisfy (2.14) one should have $c+{\dvp}^2 \approx 0$.
Given the value of $c$ in (2.1) (i.e. $c=16$ for $D=2 , \   \a'=1$)
this implies that the RG parameter $\tau$ should be identified with
with the euclidean time $\tau =i t$.

We shall return to the analysis of eqs.(2.14)--(2.18) in Sect. 3.

3.  Let us now consider the case when the matter action is represented by
the quantum vacuum energy of the string modes, i.e. by the
partition function of the string compactified on a torus. In
general, the partition function has the following duality invariant
structure
$$ Z(\l , \vp ) = \sum_{n=1}^{\infty} \e{2(n-1)\phi } \ Z_n (\l )=
\sum_{n=1}^{\infty}\e{(n-1)\vp} \ f_n(\l ) \ \ , \ \eq{2.29} $$
$$ Z(\l_i \ , \vp) = Z(- \l_i \ , \vp) \ , \ \ f_n(\l_i \ , \vp ) =
f_n(- \l_i \ , \vp )    \ \ . \eq{2.30}   $$
In the large radii limit
$$ Z_n (\l_i \rightarrow \infty ) = d_n  V   \ , \ \ V \equiv  \exp
\sm \l_i \ \ , \ \eq{2.31} $$
i.e.  $ Z$ is proportional to the volume factor. Comparing with
field theory and assuming that string  ultraviolet cutoff corresponds
to a proper time cutoff  one  may  say that it is the momentum mode
contribution
that controls the large radius limit.  In that sense
 the contributions of
the winding modes (that make  $ Z$ symmetric under (2.6) ) control the
small radii limit
$$ f_n (\l_i \rightarrow - \infty ) = d_n \exp ( - n \sm \l_i ) \
, \ \ f_1 (\l_i \rightarrow - \infty ) = d_1 V^{-1} \ . \
\eq{2.32} $$
This interpretation  is,  in fact, ambiguous since from the cutoff field
theory point of view both momentum and winding modes contribute similar
terms ( e.g. with positive and negative powers of radii) to the
partition function.
 The temperature dependence can be included by compactifying the
euclidean time [21]  (so that the inverse temperature $\b$ will couple
to $\sqrt { -  G_{00} } $ ) and does not change
the duality invariance property of $Z$ .
 Assuming adiabaticity of
evolution we may replace $\l_i ,\vp$ and $\b$ by functions of time
and take the matter action in the form (see also [22])
$$ S_m = \int dt \sqrt {- G_{00}}  \ F ( \l , \vp , \b  \sqrt { -
G_{00} }  ) \ \ , \ \ \ \ \ F = V {\cal F }\ \ , \ \  Z=-\b F \
 \ . \eq{2.33} $$

Let us consider, for example, the one-loop contribution when $F$ does
not depend on $\vp$    ( this is
a good approximation if the effective
coupling    $\rm e^{\vp}$   is small). The energy-momentum tensor can be
 represented
in the form
$$ t_{\mu \nu} = {2 \over \sqG } { \delta S_m \over \delta G^{\mu
\nu }} = {\rm diag } ( \rho\  ,\  \e{2\l_1} p_1\  , ... , \ \e{2\l_N } p_N )
\ \ , \ \eq{2.34} $$
$$ \rho= -{2\over V } {\delta S_m \over \delta \Goo } ={ E \over V } \
\ , \ \ \ \  E= F + \b {\pa F \over \pa \b} \ , \eq{2.35} $$
$$ p_i = - {1\over V } { \delta S_m \over \delta \l_i } ={ P_i \over V} \
 \ , \ \ \ \ \ P_i = - {\pa F \over \pa \l_i } \ \ . \eq{2.36} $$
 The conservation of the energy (following from the
invariance under reparametrisations of time ) implies
($\dot E \equiv dE/dt$ )
$$ \dot \rho + \sm \dl_i ( \rho + p_i ) = 0 \ \ , \
\ \ \  \dot E + \sm \dl_i P_i = 0 \  \ . \eq{2.37}  $$
 Since
$F=F(\l (t) , \b (t) ) \  $  eq.(2.37) is equivalent
to the
conservation of the entropy ${\cal S} = \b^2 \pa F/ {\pa \b} $, $\ \
\dot E + \sm \dl_i P_i = {\dot {\cal S }} / \b $ .

The resulting cosmological
equations  (2.9) -- (2.11) are [12]
$$  c - \sm \dl_{i}^2  + \dvp^2   =  {  \evp} E  \ \ , \ \eq{2.38}  $$
$$ \ddl_i - \dvp \dl_i  =   \half  {\evp }  P_i \ \ , \eq{2.39}  $$
$$ \ddvp_i - \sm \dl_{i}^2 =  \half  {\evp } E   \ \ . \eq{2.40}  $$
The time derivative of (2.38) vanishes on eqs.(2.39), (2.40) if
eq.(2.37) is satisfied. Eqs.(2.37)--(2.40) are invariant under the
duality transformation (2.6)  if $ F$
 is duality symmetric (note that
under the duality $ E \rightarrow E \ , \ \
P_i \rightarrow -P_i$
).

Solving the adiabaticity condition one can in principle express the
temperature in terms of $\l_i$ . Then eqs. (2.38)--(2.40) will take the same
form as in the zero temperature case  with $E$ and $ P_i$ replaced by
$\CE (\l) $ and $ - {\pa \CE }/ {\pa {\lambda_i} } $.

Re-written in terms of the original dilaton field $\p$  (see (2.5))
eqs.(2.38)--(2.40) take the form
$$ c- \sm \dlis + ( 2\dop - \sm \dl_i )^2 = \e{ 2\p } \ {\r } \  \  ,
\eq{2.41} $$
$$ {\ddot \l}_j - (2\dop - \sm \dl_i ) \ \dl_j =  \half
 \e{2\p} \ { p_j}  \ \ , \eq{2.42} $$
$$ 2\ddp - \sm  \ddl_i - \sm \dlis = \half  \e{2\p} \ {\r}  \ \ .
\eq{2.43} $$
It is easy to see that if $c=0$ a solution with
  $\p = \rm const \ $  may exist only if the energy-momentum tensor (2.34) is
  traceless, i.e.
$$\sm p_i = \r \ \ . \eq{2.44}  $$
 The latter relation ( which is not duality
symmetric)  is not in general
satisfied  by the  free energy which actually appears in string theory.
This point was already emphasized in [12] (cf. [3,4]).

A simple way to see that $\p= $ const is not in general
 a solution of the cosmological
equations if the one-loop vacuum energy is used as a source is to start
with the action rewritten in terms of  the ``rescaled'' metric (cf.(2.1),(2.33)
)
$$  G'_{\mu \nu } = \e{-4\p /(D-2)} \Gmn \ , \ \  \eq{2.45} $$
$$ S= - \int d^D x  \sqG \ \e{-2\p} \ [ \ c + \ R  + \ 4 (\pa \p)^2 ] \ + \
  \int d^D x \sqG \ {\cal F } =  $$
$$ - \int d^D x {\sqrt {-G'}} \ [ \ R' - {4\over D-2 } (\pa \p)^2 + c\
 \e{4\p/(D-2)} \ ] \ +
 \ \int d^D x {\sqrt {- G'}} \ {\rm e }^{2D\p / (D-2)}
  \ {\cal F} \ . \eq{2.46} $$
The equation for $\p$ which follows from (5.10) is not solved by $\p=$const
if  $c=0$ and $F\not=0$.

4. Before turning to the analysis of eqs. (2.9)--(2.11) with
duality symmetric matter  let us recall  their solution in the
absence of matter [15].  If $S_m = 0$  eq.(2.10) implies
$$ \dl_i = k_i \ {\evp } \ \ , \ \ k_i = \rm const   \ \ .
$$
Substituting this into eq.(2.9)  we finally get ( assuming $D >
26$ )
$$ \vp = \vp_0 - {\ln  \sh 2} bt   \ \ ,\ \  \l_i = \l_{i 0} +{
q_i}\  {\ln \th }bt \ \ , \eq{2.47} $$
$$ \sm q_{i }^2 =1 \ \ , \ \ \ \a' b^2 = {1\over 6 } (D-26) \ \ , \ \
\ q_i = k_i b^{-1} \e{\vp_0}  \ \ . \  $$
The solution for $ D < 26$ is obtained by the substitution $b
\rightarrow  ib $ ;  in particular , for $ D=2$
$$ \vp = \vp_0 - {\ln \  \rm sin} \ 2 bt   \ \ , \ \ \l =\l_0  \pm
{\rm ln \   tg \ } bt \ \ , \ \ \a' b^2 = 4  \ . \eq{2.48} $$
 The solution for the zero central charge ( $D=26$ ) is given by
$$\vp = \vp_0 -  {\ln } \ t  \ \ , \ \ \l_i =\l_{i 0} + q_i \ { \ln }\ t \ \ ,
\
\ \sm q_{i}^2 =1 \ . \eq{2.49} $$
 This is a Kasner-type solution with some radii infinitely
expanding and others infinitely contracting. Such behaviour will be
changed by the presence of matter. Note that the duality
transformation  (2.6) corresponds to $q_i \rightarrow - q_i $  ,
i.e. it relates contracting solutions to expanding ones.

\newsec{Solutions with classical matter}

1.  In this section we shall study the solutions of the system of
equations (2.14)--(2.18). We shall make a number of simplifying
assumptions.
In what follows we shall consider  the
isotropic case of all time dependent radii being equal $\l_i = \l$.
This assumption is  consistent with eq.(2.15) if the modes $\ps$
and $\tps$  are such that $m_i=m \ , \  \tilde m_i =
\tilde m \ , \ i=1,...,N $.
The coresponding metric (2.4)
$$ ds^2 = - d t^2  +  \e{2\l (t) } \sm dx_{i}^2    \ \ , \
\eq{3.1} $$
can be also interpreted as a spacially flat isotropic cosmological
metric. Then
$$ \vp = 2 \p - N \l \ \ ,\ \ \ \   V  = \e{N\l} \ \ . \eq{3.2} $$
To be able to change the value of the central charge we shall
assume that there may be a number $  K$  of ``static'' spacial
dimensions, i.e.
 $$ D = 1 + N + K \ \  ,\ \ \ \  c= -{2\over 3} ( N+K-25) \ \ .
$$
 In general, the duality
transformation (2.6)
$$ \tl = - \l \ \ , \ \ \ \tvp = \vp \ \  \eq{3.3}  $$
corresponds to  inverting the direction of evolution of the scale factor.

 Eqs. (2.14)--(2.18) reduce to
$$ c- N\dl^2 +\dvp^2 = \half \ [\ \dpss + ( Nm^2 \e{-2\l} - 4 ) \pss
$$
$$ +\  \dtpss + ( N \tm^2 \e{2\l} - 4 ) \tpss  + 2 U \ ] \ , \eq{3.4} $$
$$ \ddl - \dvp \dl = \half \ (\  m^2 \e{-2\l} \pss - \tm^2 \e{2\l} \tpss
\ ) \ , \eq{3.5} $$
$$ \ddvp - N \dl^2 = \half \ ( \ \dpss + \dtpss \ ) \ , \eq{3.6} $$
$$ \ddps - \dvp \dps + (Nm^2 \e{-2\l} -4 ) \ps  + {\pa U \over \pa \sps}
=0 \ , $$   $$  \ \
\ddtps - \dvp \dtps  + ( N\tm^2 \e{2\l} -4 ) \tps
+ {\pa U \over \pa \stps }= 0 \ . \eq{3.7}
$$

We shall first ignore the dynamics of the matter fields $\ps$ and
$\tps$  and consider them as  ``static'' sources. Such an assumption
may be justified
  if the the full interaction potential ( including $U$ as well as
interactions of $\ps $ and $\tps$ with other modes which we ignored)
 has extrema (see, however, the discussion at the end of this section).
 Then  eqs.
(3.4)--(3.6) take the form
$$ -N\dls + \dvps = 2 N W (\l ) \ \ , \ \
 \ \ W \equiv  \half \mu \ \e{-2\l} + \half  \tmu \  \e{2\l} - C/2N
  \ , \eq{3.8} $$
$$\ddl- \dvp \dl = \mu \ \e{-2\l} - \tmu \ \e{2\l} = - W'(\l ) \ ,
\eq{3.9} $$
$$ \ddvp - N \dls = 0 \ , \eq{3.10} $$
where
$$ \mu= \half m^2 \pss \ \ , \ \ \tmu =\half \tm^2  \tpss \ \ , \ \ C \equiv {
c
+ 2 \pss + 2 \tpss  - U (\ps , \tps )  } \ \ . \eq{3.11} $$
Note that the derivative of eq.(3.8) is still vanishing if (3.10) and
(3.11)   are satisfied.  Let us discuss some general properties of
eqs. (3.8)--(3.11) assuming that $c\ ,\  m$ and $\tm$ are such that $W$ is
strictly positive. Then eqs.(3.8), (3.10)
imply that $\dvp \not= 0 $  and  is always growing.
Hence if we start with $\dvp < 0 $ (i.e. the effective coupling
$\rm e^\vp$ decreasing with time ) then $\dvp$ will increase towards
positive values but will never reach zero. Eq.(3.9) then describes
a motion of a particle in a potential $W > 0 $ (which grows at
large positive and negative $\l$)  with a damping term proportional
to $-\dvp$ [12]. Introducing the energy of the particle $\CE$  we find
from (3.8)--(3.10)
$$ \dot { \CE } = \dvp \dls \ < 0  \ \ ,  \ \ \ \ \dvps= 2N { \CE } \ > 0
    \ , \eq{3.12}
$$
$$ {\CE } \equiv \half \dls + W ( \l )  \ , $$
i.e. $\CE$ will decrease with time. The particle coordinate $\l (t)
$  will thus be oscillating  between its maximal positive and negative
values with the amplitude of oscillations decreasing to zero (i.e.
the particle trajectory be reflecting
from the walls of the potential  moving down towards its bottom).
  If at the initial moment the scale factor $a=\rm e^{\l}$ is
increasing ($\dl > 0  ) $ the expansion will continue untill a
maximal value of $a$ is reached.  After the turning point
 the contraction will start
untill a minimal value of $a$ is reached , etc.

The asymptotic values of $\l$ and $\vp$  correspond to the minimum of the
potential $W$
$$ \l_* = {\textstyle {1\over 4}} {\ln} \ (
\mu / \tmu ) \ , \ \  \ \ \vp_* = \vp_0
- Q t \ , $$
$$ \  \  Q \equiv [ 2N W (\l_* ) ]^{1/2} = [ \ 2N(\mu \tmu
)^{1/2} - C  \ ]^{1/2} \  \ , \ \ W(\l_*) =0 \  .  \eq{3.13} $$
If $m=\tm \ , \ \ps= \tps \  $  then $ \ \l_* =0 $ and duality is
asymptotically restored  at $t \rightarrow \infty $. To find the
behaviour of the solution at large $t$  we can approximate $W$
near $\l=\l_*$ by
the oscillator potential
$$ W= A + \half  B \k^2  \ \ , \ \ \l=\l_* + \k (t) \ , \ \ A= Q^2/2N \ , \
\ B= (Q^2 + C )/N = 2 (\mu \tmu )^{1/2}  \ ,  $$
$$ \ddot \k + Q \dot \k + B \k = 0  \ , \ \ \ \k \sim \e{\o t } \ ,\  \
\ \o = - \half ( \  Q  \pm \ [ Q^2 - 4 B ]^{1/2} \ ) \ \ . \eq{3.14}
$$
Depending on the sign of $\  Q^2 -4 B= 2 (N-4) (
\mu \tmu )^{1/2} - C  \  $ we get either exponential
falloff or  oscillations with exponentially decreasing amplitude.

The solution in the case $\dvp > 0 $ can be obtained by time
reversal (see (3.8)--(3.10)). We conclude that $ \rm e^\vp$ will
be increasing  and  $\l$ will
oscillate    with the  amplitude of oscillations  exponentially
growing with time.

One can also analyse the  solutions of (3.8)--(3.10)
by studying the behaviour of the trajectories on the $(\l \ ,
\ \dvp )$ plane. We have
$$ \dot f = N \dls \ , \ \ \  \dls = {1\over N}( \  f^2 - 2 N W \ ) \ ,
  \ \ \ f \equiv \dvp \ \ ,     \eq{3.15} $$
$$ f' \equiv { df \over d \l} = \pm \sqrt {( f^2 - 2NW )/N } \ . \eq{3.16} $$
(near the tuning point $\dl =0$   eq.(3.10) should be used as well ).
The trajectories on the $(\l , f ) $ plane
will lie above and below the limiting curves
$f= \pm \sqrt {2NW(\l ) } $. Consider, for example  the case when
$f > 0  \ ,  \l >0 \ , \ \dl > 0 $ . It is easy to see that the
trajectory will move up (
$\dot f  > 0
$ )  towards the right branch of the  limiting curve and will hit it in a
finite time.
    We shall present  the explicit  solution of
eq.(3.15) in the region of large $\l$
(and for $N=1$ ) in Appendix A where we shall demonstrate
that  the growing $\l (t) $ always reaches a maximal value.
    Since at the turning point $\dot f =0 \ , \ \dot \l
=0 $  but $\ddl < 0 $  the trajectory will be reflected back : $\l$
will start decreasing , the trajectory will go up until it will
finally hit the left branch of the limiting curve ( $a$ will reach
its minimal value). If we start with $f < 0 $ the trajectory will
lie below the lower limiting curve.  $f $ will increase
along the trajectory  so the amplitude of oscillations of $\l$ will
be decreasing asymptotically to zero. Again , the two cases ( $f >
0 $ and $ f< 0 $ )  are
related by time reversal.

2.   The above qualitative analysis of the system (3.8)--(3.10) can
be repeated for the general system (3.4)--(3.7) with time dependent
matter fields. Since according to (3.6) $\ddvp $ is again non-negative
$\dvp$ will remain negative if we start with decreasing $\vp$. Then
  the damping terms in the equations for the matter fields
(3.7) will cause  $\ps  \ ,\  \tps $ to decrease with time towards their
``background'' ( presumably constant) values.  The ``potential'' $W$ in
(3.8) will now include the positive
kinetic terms for $\ps$ and $\tps$ and hence the
energy $\CE$ in (3.12) will still satisfy $ \dot {\cal E }  < 0 $.
Since the derivatives of the matter fields will decrease with time
the qualitative behaviour of the solution for $\l$ ( oscillations
with decreasing amplitude ) will remain the same.\foot { If the
structure of matter interactions is such that there is no
non-vanishing backgrounds for $\ps$ and $\tps$ then the matter
field contributions  to eqs.(3.4) -- (3.7) will be  decreasing to
zero.  Hence the late time solution will be approaching   the
``free'' solution (2.47) and the period of oscillations will become
infinite.}

To summarise,  if we start with the initial condition $\dvp < 0 $
,  i.e. the effective coupling $ \rm e^\vp $ decreasing with time (
this is a natural assumption in order to justify the neglect of
quantum corrections in the matter action ) then the scale
 $a$ oscillates between the maximal and minimal values.
If the matter action is ``self- dual'', i.e. $\mu =\tmu$  then
 $a_{max} $ and $ a_{min} $ asymptotically approach the
Planck scale $\sqrt {\a'} =1 $ , i.e. $\l=0$. If one starts with
 a large value of $a$ it will take, of course, a large number of
oscillations  before $a_{max}$ will become of Planck order.

As it is clear from the structure of the potential $W$
 the role of the winding mode contribution $( \sim \exp 2\l )$ to
the ``classical'' system of equations (3.4)--(3.6) is to
prevent infinite expansion . The winding mode term in eq.
(3.4) is thus {\it opposing } the expansion [12]. Without it
we would have either the  expansion to $\l = \infty$ or first
contraction to some minimal radius and then infinite expansion.
The role of the momentum mode contribution is ``dual'': it
prevents contraction to $\l = - \infty$ , i.e. provides the
existence of a non-zero minimal radius of contraction.

One may question the validity of the assumption  used in deriving
 (3.8)--(3.10) that the matter fields
in (3.4)--(3.7) may be taken  to be constant. It may be more appropriate
to set  constant the rescaled fields $ \ps'= \exp (-\vp /2 )  \ps \ , \
\tps' = \exp (-\vp / 2 ) \tps \  $   for which there is no $\dvp$ term in the
analog of eq.(3.7). Then  eqs.(3.8)--(3.9) will have the same form but with
extra factor of $\exp {\vp}$ appearing in front of $W$ (eq.(3.10) will be
 modified
accordingly to preserve the consistency of the system (3.8)--(3.10) ; being
 redundant,
it may be ignored). Similar system appears in the one-loop ``quantum'' case
discussed in the next section and the behaviour of the solution is similar.
Let us note that the presence of the $\exp {\vp}$ factor in the r.h.s. of
 equations
corresponding to the classical  string modes is suggested also by the form
of the  combined action (2.7) where $S_m$ is given by the the classical
string  ($\s$ model) action.
The absence of the dilaton factor in front of the classical string action
implies its presence in the r.h.s. of the corresponding cosmological equations
[10]. Hence a condensate of classical string modes is probably described by the
rescaled classical fields. The presence of the dilaton factor in the r.h.s.
of equations describing the cosmological evolution
in the case when the ``matter'' is represented by  classical
string modes  was also assumed in [12] where the conclusion that the winding
modes oppose the expansion was already made.

\newsec{ Solutions with quantum matter }

1. In this section we shall study
the cosmological evolution in the case when the matter action in
(2.2)--(2.3) is represented by the (zero temperature)
 one loop partition function of
the string  compactified on the torus. The general
form of the solution will be the same as in the case
of the classical matter discussed in Sect. 3
: the scale factor will oscillate between its maximal and minimal
values with the amplitude of oscillations decreasing to zero or
increasing to infinity depending on whether the dilaton coupling is
decreasing or increasing with time.

As in Sect.3 we shall consider the isotropic case,
$$ \l_i =\l \ \ , \ \ F(\l_1 \ ,... , \l_N ) = F(\l ) \ , \ \
P_i= - {\pa F \over \pa \l_i } = - {1\over N } {\pa F\over \pa
\l} \equiv P \ . \eq{4.1} $$
Then eqs.(2.38)--(2.40), (2.37) take the form
$$ c-N\dls +\dvps = \e{\vp} E \ \ , \eq{4.2} $$
$$ \ddl -\dvp \dl = \half \e{\vp} P \ \ , \eq{4.3} $$
$$ \ddvp - N \dls = \half \e{\vp} E  \ \ , \eq{4.4} $$
$$ \dot E + N \dl P =0   \ \ . \eq{4.5}   $$

2. For completeness let us first  discuss the case of a non-zero temperature
   ($ F= F( \l (t) , \b (t) ) $ )  and
 consider (4.2)--(4.5) as a ``phenomenological''
system of equations which should be supplemented by some equation
of state relating   $E$ and $P$   ( for more general
analysis of the finite temperature case  see [12]).
In the case of the usual radiation -- type condition
$$p=\g \r \ \ ,\ \ \ \ P=\g E   \ \ ,\eq{4.6} $$
 we find from (4.5)
$$  E= E_0 \exp (-N\g \l ) \ \ , \ \ P= \g E_0 \exp (- N \g \l ) \
\ . \eq{4.7} $$
Then a particular solution of (4.2)--(4.4) with $c=0$ is given by ( cf.
(2.49) ; similar solutions were discussed in [22,11] )
$$ \vp = \vp_0 + s \ {\ln }\ t \ \ , \ \ \l = \l_0 + q \ {\ln} \ t  \ \
, \eq{4.8} $$
$$ s=-2 / (1+ \g^2 N ) \ , \ \ q= 2\g / ( 1+ \g^2 N ) \ \ ,
\eq{4.9} $$
$$ E_0 =  4 (1-\g^2 N ) ( 1+ \g^2 N )^{-2}\  {\exp (N\g \l_0 - \vp_0 ) }\
\ , \ \  s-\g N q = -2  \ \ . $$
 One should assume $\g^2 N < 1 \  $ to have $E_0>0$ . If  $\ \g = 1/N $
(so that (2.44) is satisfied)
$$ s=- 2N / (1+N )   \ ,\ \ q = 2 /(1+N) \ , \ \ \p= \half (\vp + N \l
) = \rm const  \ \ . \eq{4.10} $$
This solution  describes a power -- law  expansion with the dilaton
coupling decreasing with time.

The above equation of state (4.6) explicitly breaks the duality
invariance of (4.2)--(4.5). As we shall see, the behaviour of the
solution is very different in the case when the duality invariance
is preserved. The simplest  duality invariant equation of state is
$P=0$ , i.e. $E=E_0 =\rm const$. This seems to be a natural choice
for an equation of state near the Hagedorn temperature [2]. In
general ,  if the temperature is approximately constant and $F$ is
duality symmetric such a relation will always
be true near the Planck scale $\l =0$, i.e.  $\ E(\l) = E(-\l) \approx
{\rm const} \ , \ P = - F'/N \approx 0 $ . In that case we find
from (4.2)--(4.4)
$$ \dl = A \e{\vp} \ \ , \ \ \dvps = NA^2 \e{2\vp} + E_0 \e{\vp} -
c \ \ . \eq{4.11} $$
For $c=0$ [12]
$$ \vp = \ {\ln \ } \vert  {4E_0 \over {E_0}^2
t^2 - 4 N A^2  } \vert \ \ , \ \
\l=\l_0 + {1\over \sqrt N }   \ {\ln } \ \vert {{t- 2{\sqrt N} A/E_0 }
\over { t+ 2 {\sqrt N} A/E_0 }} \vert \ \ , \eq{4.12} $$
i.e. we conclude that $\rm e^{\l}$ is asymtotically approaching its maximal
or minimal values in the case of expansion ( $A>0$ ) or contraction
($A<0$ ) while the dilaton is decreasing with time. The duality
transformation corresponds to $A \ra -A$   or, equivalently,
 $t \ra -t$  . Note that
$\vp=\rm const$ is not a solution of (4.2)--(4.4) (and $\p \not= 0
$ for $\g=0$ , cf.(4.10) ).

3. Let us now turn to the analysis of the solutions of
eqs.(4.2)--(4.5)  for the duality symmetric zero temperature case,
i.e.
$$ E = F(\l ) =F(-\l)
 \ , \ \  \ P =-F'/N \ , \ \ \ F( \l \ra \pm \inf )  \ra  d_1
\e{\pm N\l} \ . \eq{4.13}   $$
Let us  first formally assume that $E$ is positive (we shall
consider the case of negative $E$ later on). Though the Casimir
energy in field theory is  usually negative, the bosonic string vacuum energy
is
 in any case
not well defined (divergent due to tachyon)  but the aim of our analysis
is only to illustrate some general properties of the system (4.2)--(4.4).

Assuming (4.13) one finds that eq.(4.5) is satisfied automatically
and (4.2)--(4.4) take the form
$$ -N\dls +\dvps = 2 N W \ ,\  \ W(\l,\vp)=  [ \e{\vp}
F(\l ) -  c  \ ]/2N \ , \ \ \eq{4.14} $$
$$ \ddl - \dvp \dl = - {\pa W \over \pa \l} \ , \ \ \eq{4.15} $$
$$ \ddvp - N\dls = N W - \half c  \ \ . \ \eq{4.16} $$
Since $F(\l) $  ( and  hence $W(\l)$ )  has the minimum at $\l=0$
there exists the ``fixed point''  self-dual solution
$$ \l = 0 \ \ , \ \   \dvps = F_0 \e{\vp}  - c \ \ ,  \ \ F_0 = F(0) > 0
\ , \eq{4.17} $$
i.e. if $c > 0$ and $ \dvp < 0 $
$$ \vp= \vp_0 - 2 {\ \ln }\ | {\sin \ } \half{\sqrt c }\ t | \ \ , \ \
\ \ \e{\vp_0} = c/F_0 \ . \eq{4.18} $$
Solutions with generic initial conditions will ``spontaneously break''
duality.

  A qualitative analysis of a general solution of (4.14)--(4.16) is
similar to that of the ``classical system'' (3.8)--(3.10).
Let us consider the case of $\dvp < 0 $.
We note that since  $\ddvp $  is positive $\dvp$ will grow but will
remain negative if it was negative at the initial moment, i.e.
$\vp$ will be monotonically decreasing. Treating
$\vp (t)$ as a given function  we can then interpret
 eq.(4.15) as the equation of motion for a particle with the  damping
term $ \sim -\dvp$ and the time dependent potential $ W$. The
``energy'' of the particle is decreasing with time (cf.(3.12))
$$  {\CE} \equiv \half \dls + W \ , \ \ \ \dot {\CE } =  \dvp ( \dls +
\e{\vp} F/2N  ) \ < 0 \ , \ \ \dvps= 2N {\CE } > 0  \ .  \eq{4.19} $$
If we draw the graph of the potential $W(\l , t )$ (which is
exponentially rising at $\l \ra \pm \inf $ and has a minimum at
$\l =0$ )   at several consequent
moments of time  it will be moving down towards the $\l$ axis. We
shall assume that $c$ is such that $W > 0$ . Since $\CE$ is
decreasing the trajectory of the particle will also be moving down
reflecting from the walls of the potential. As a result, the radius
$\rm e^\l$ will be oscillating from  maximal to minimal values
with the amplitude of oscillations asymptotically decreasing to
zero, i.e. the solution  will be
approaching the self-dual point
,  $\l (t \ra \infty) = 0 $.  The solution in the region
near $\l =0 $ where $ F \approx \rm const $ was already analyzed
above (see (4.11), (4.12)) :  $\l$ is asymptotically approaching
its constant value (zero) while $\rm e^\vp$ is decreasing to zero.

If we start with $\dvp > 0$ the direction of  evolution is
reversed: the amplitude of oscillations  of $\l$ grows  with time,
i.e. the maximal and minimal values of $\l$ are asymptotically
approaching $\pm \inf$ .

  It is possible to relate the analysis of the system
(4.14)--(4.16)  to that of the system (3.15)--(3.16) which appeared
in the ``classical'' case. Introducing the new time coordinate
$d\tau = dt {\rm e}^{ \vp / 2 } $ and assuming $c=0$ we find from
(4.14)--(4.16)
$$ N\l'^2 = f^2 - F(\l) \ , \ \ 2 f' = f^2 - F(\l) \ , \ \ f\equiv
\vp' = {d\vp \over d \tau } \ ,  \eq{4.20} $$
$$ { d f \over d \l } = \half {\sqrt N [ f^2 - F ( \l ) ]^{1/2} }
  \ . \eq{4.21 } $$
Eq.(4.21) is similar to eq.(3.16) and so is the behaviour of the
corresponding trajectories on the $(\l \ , \ f )$ plane.

In Appendix B we shall present the explicit solution of
eqs.(4.14)--(4.16)  in the asymptotic region of large $\l$ which
demonstrates that a  finite extremum value of $\l$ is always
reached in a finite time ( i.e. any trajectory  on the
$(\l ,f)$ plane always hits a boundary curve ).

Let us now consider the case when $E$ in (4.13) is negative.
Then it follows from (4.14) that $\dot \l $ can never vanish
 (we are assuming that $c$ is non-negative). As a result, there
 can be no turning  points, i.e. the expansion goes to infinity
 and contraction goes to zero.

To illustrate what happens  when $W$ is negative
let us consider the special case
of $N=1$.  If there are no
extra static space dimensions, i.e. $D=2 \ , \ c=16 \  $
then we may use the  expression for the vacuum energy (2.29), (2.30)
of a $D=2$ string compactified on a circle
which is known to all loop orders [23]
$$ f_1 =d_1 ( \e{\l } + \e{-\l }
) \ , \ \ f_2 = d_2 ( \e{2\l } + \e{-2\l } +
10 / 7 ) \ ,  \ \  ... \ \ , \ \   \eq{4.22} $$
where $d_1$ is negative (see also [12]).
Assuming again that the effective coupling $  {\rm e^{\vp}} $ is
sufficiently small
so that we can consider  only the one-loop term  $ F=f_1$
let us describe another  way of solving (4.14)--(4.16) for $N=1$
in the large radius limit.
If $\l$ is large   eqs.(4.14)--(4.16) can be represented in
the form
$$ \ddot \r - \dl \dr =0 \ ,\ \ \dr (\dr - 2 \dl ) = d_1 \e{\r} - c \ , \ \
\r \equiv \vp + \l \ , \eq{4.23} $$
i.e.
$$ \dr = B \e{\l} \ , \ \ -2 \ddot \r + \dr^2 = d_1 \e{\r} - c \ .
\eq{4.24} $$
Integrating once the equation for $\r$ ( introducing $ y= \exp (-\ha
\r) $ ) we find
$$ \dr^2 = q\  \e{\r} - d_1 \r \ \e{\r}
- c  \ \ , \ \ \  q= \rm const \ \ .  \eq{4.25} $$
As a result, if the radius ( which according to (4.24)
is proportional to $\dr$ ) is large and growing,
 it  grows indefinitely.

\newsec{ Concluding remarks }

1.  It useful to clarify the meaning of the duality invariance (2.19),
(3.3) of our
systems of equations (3.4)--(3.7) and
(4.14)--(4.16).  Let us  consider, for example, (3.4)--(3.7) and assume
for simplicity that  $m=\tilde m$.
Then the duality invariance implies that for each solution $(\l\
  , \ \vp\ , \ \ps \ , \ \tps \ )$ there exists another solution
$(\ - \l \ , \ \vp \ , \ \tps\ , \ \ps \ )$.   If we
consider the simplified system (3.8)--(3.10) and further assume
that $\mu = \tmu $  then the potential in (3.8)
 $W= 2\mu \cosh 2\l - C/2N $   will (like the potential in
(4.14)) have the
exact symmetry under $\l \rightarrow -\l $.  Though there will exist
the symmetric (``self-dual'') solution (cf.(3.13), (4.17), (4.18))
 $ \ \l=0 \ , \ \vp =\vp_0 - Qt \ $  in general duality will be
``spontaneously broken'' (by the initial condition
$ \l (0) \not= 0 $)  on  a time--dependent solution. As we have
seen, however, the duality is asymptotically restored at late times
(for a generic choice of initial conditions for $\l \ , \ \dl \ $ and
$\dvp < 0 $ ) : the oscillating solution $\l (t) \rightarrow
0 $  at $ t\ra \infty $.
For all the oscillating
 solutions the duality is true in the following ``average'' sense : the
average value of $\l   (t)$ is zero. Also,  the behaviour of a  given
solution at large positive and large negative $\l$ is similar .
This is in agreement with one's expectation that duality should be a
symmetry between processes at  large and small distances.

2. Let us now discuss a
model in which some of time
dependent space dimensions are ``uncompactified'', i.e. the vacuum
energy   depends on their scale factor in a trivial  way. Let us
consider the zero temperature case and set $c=0$. Denoting the common scale
of compact (noncompact) dimensions by $\l$ ($\L$) and their number by
$N$ ($n$) we find from
eqs.(2.38)--(2.40)
$$ - n\dL^2 - N \dl^2 + \dvps = \e{\vp +n\L} F \ , \ \ \
F=F(\l)  \ , \ \ \eq{5.1} $$
$$  \ddL - \dvp \dL = -\half \e{\vp + n \L } F \ , \ \ \eq{5.2} $$
$$ \ddl -\dvp \dl = - \half  N^{-1} \e{\vp + n\L } F'   \ , \eq{5.3} $$
$$ \ddvp - n\dL^2 - N\dls = \half \e{\vp +  n \L } F  \ . \eq{5.4} $$
Assuming $F$ is positive  a
qualitative analysis of this system is similar to that of
eqs.(4.14)--(4.16). Since $\ddvp > 0$ the  dilaton continues to
decrease monotonically , providing the damping terms in the equations for
$\L$ and $\l$ . Combining eqs.(5.1) , (5.2) and (5.3) one finds (cf.
(B.1),(B.2))
$$ \ddL -\ddvp = \dvp (\dL-\dvp) \ , \ \ \ \dL= \dvp + A \e{\vp} \ . \eq{5.5}
$$
As a result, one  concludes that $\L$ first expands to its maximal value and
 then contracts to
zero (note that there is no winding mode contribution which may protect
$\L$ from contraction to zero)
 while $\l$ oscillates between maximal and minimal values  approaching
the self-dual point $\l= 0 $.

 Since $\L$ in (5.1)--(5.4) corresponds to the non-compact dimensions it is
more natural to introduce
 another ``shifted'' dilaton field  as an effective coupling (cf. (2.5))
$$ \bvp \equiv 2\p - N \l \ \ , \ \ \ \vp = \bvp - n\L  \ \ . \eq{5.6} $$
 $\bvp$ ``absorbes'' only the volume factor
corresponding to compact dimensions  and like $\vp$ is invariant under the
duality $\l \ra -\l$ . The weak coupling regime corresponds to $\bvp$
decreasing with time. $\bvp$ in general will not be monotonic
but will be decreasing at late times (if $\dvp <0$).
   Let us  consider the special solution
with constant ``internal'' scale
 $\l=0$, i.e. $\l$ sitting at the minimum of $F$. Then (5.1),(5.2) and
(5.4) become identical to the system (B.1) (with $N,\l,d_1 $ replaced by
$n, \L, F(0)$ ) the solution of which is given by
 (B.8),(B.9),(B.12). The behaviour of $\bvp$
$$\bvp= \vp + n\L = \bvp_0 + (\sqrt n +1)/(\sqrt n -1)\ {\ln}\ z -
2(n+1)/(n-1)\ {\ln} (1+z)  \  \eq{5.7} $$
is similar to that of $\L$ : it first grows with $t$,  reaches its maximum
(at $z= (n+1)^2 /(n-1)^2 $) and then starts decreasing. Since $\L$ reaches its
maximum at a later time $t$ ($z=(n+1)/(n-1)$) there exists the interval of $t$
in which the non-compact dimensions
are still expanding while the dilaton coupling is already
decreasing.

3. Cosmological solutions for the space being a
 product of a number of large flat dimensions and an ``internal'' torus
 (and a finite temperature  one--loop vacuum energy  as a source)  were
previously discussed in refs.[3,4] with a conclusion
that the presence of the winding modes prevents internal dimensions from
expanding, stabilizing them at late times near the Planck scale.
 The dilaton dynamics was, however, ignored,
 i.e. the original dilaton $\p$ (see (2.5)) was implicitly
 assumed to be
constant.  We would like to emphasize that
this assumption is inconsistent in general:
as was already noted in Sect.2
 the system (5.1)--(5.4) does not have solutions with
 $$\p \equiv \half ( \vp + n\L  + N \l )  = \rm const \   $$
unless  (2.44) is true. The latter condition can be satisfied only in the
limiting cases when the contributions of the winding modes can be ignored.
\foot {
Let us note also that no solution of (5.1)--(5.4) exists with $\bvp= \rm const$
(the substitution of $\dvp=-N\dL \ $ into (5.1)--(5.4) gives $ F=0$ ).}

In this paper we considered only models at zero temperature.
As we have already mentioned the more ``realistic''
finite temperature  case was  discussed in [12].

\vskip .4in

 I would like to acknowledge J. Bagger and  M. Tsypin for help
and C. Vafa for collaboration on ref.[12] on which
most
of the present paper
is based. I am also grateful to Trinity College, Cambridge for
a financial support.

\vskip .3in

\appendix{ A } { Asymptotic solution of the system (3.8)--(3.10) }

Here we shall present the solution  of the ``classical'' system
(3.8)--(3.10) in the case of $N=1$.  Our aim is to demonstrate  the
correctness of the
qualitative analysis of the behaviour of the solution  given in Sect. 3
which predicts the existence of a finite maximal radius of expansion.
In order to solve eq.(3.16) explicitly we shall  consider the region of
large positive $\l$ in which the potential $W$ can be approximated by
$W= \ha \tmu \e{2\l} $ . Assuming that $\dl > 0 $ and
absorbing the positive constant $\tmu$ into $\l$
we find that eq. (3.16) takes the form
$$    f' = \sqrt { f^2 - \e{2\l} }  \ \ , \eq{A.1} $$
or, equivalently,
$$ k' = - h (k) \ , \ \ h \equiv k- \sqrt{k^2 -1} \ , \ \  \ \ k(\l ) \equiv
f(\l ) \e{-\l}  \ , \ \ k=\half ( h + h ^ {-1} ) \ \ . \eq{A.2} $$
Integrating over $h$ we get
$$ - \half { \ln } \ \vert h \vert - {1\over 4 h^2 } =  \l - \l_0 \equiv \bl
\ \ . \eq{A.3} $$
Here the integration constant $\l_0$ should be large positive. Suppose that
$f >0 $ , i.e. $ f> 1 \ , k > 1 $  so that $ h $ changes from 1 at $k=1$ to
0 at $k= \infty$. Then $\overline \l
$ increases from some large negative value at
initial large $k$ to its maximal value $-1/4$ at $k=1$.  The maximal value
of $\l$ ( or the value of $\l_0$ ) is determined by initial conditions for
$\l$ and $f$.  We have thus proved that for the given exponentially
growing potential
the rising trajectory on the $ (\l , f ) $ plane will always hit the
boundary curve so that the maximal value of $\l$ will be reached in a
finite time.

The case of $f < 0 \ , \ \ \dl > 0 $ is analysed in a similar way. Then
$h$ is strictly negative and (A.3) is still valid. Both $k$ and
$\overline \l$ increase from their initial negative values to their
maximal values $k=-1$ and $ {\overline \l}=-1/4$.

If $\dl < 0 \ \ $  at $\ t=0$   $\ \  \l \ $ will be decreasing towards
negative values and will finally reach the region where the
constant and $\rm e^{-2\l}$
terms in the potential $W$ (3.8) cannot be ignored.   The solution in
 asymptotic region of large negative $\l$  can be found
by the duality transformation: $\l \rightarrow -\l$ .  The potential will
be dominated by the term $\rm e^{-2\l}$ which will imply the
existence of a minimal radius of contraction.  The full solution
can be found by ``sewing'' the solutions in the two asymptotic
regions.

There is another way of representing the solution of (3.8)--(3.10) in the
asymptotic regions of large $\pm \l$. Consider, for example, $\l < 0 $
and $\dl > 0 \ , \ \dvp < 0 $ , i.e. the case of expansion  starting from
small radius.  Introducing the ``conformal time'' $\tau$
$$ d\tau = \e{-\l } dt \ \ , \ \ \ \ \ { F' \equiv {dF \over d\tau }} = \e{\l }
 F  \ , \eq{A.4} $$
we find from (3.8)--(3.10) $ \ (N=1)$
$$ \r'' - \r'^2 =0 \ , \ \ -\r'^2 + 2 \r'\vp' = \mu  \ , \ \ \r \equiv \l +
\vp \ , \eq{A.5}  $$
$$ \r = \r_0  - \ln \ |\bt | \  \ , \ \ \vp = \vp_0 - \half \ln \
|\bt |  - \mu  \bt^2 /4   \ , \ \ \bt \equiv \tau - \tau_0 \ ,
\eq{A.6} $$
$$ \l = \l_0 - \half \ln \ | \bt |  +  \mu \bt^2  /4 \ , \ \
dt = d\tau \tau^{-1/2} \exp ( \l_0 +  \mu \tau^2  /4 ) \ . \eq{A.7}
$$

\appendix{B}{ Asymptotic solution of the system (4.14)--(4.16)}

Below we shall solve eqs.(4.14)--(4.16) in the region where $\l$ is large and
positive ( the solution for large negative $\l$  can then be found
 by the duality transformation).
 Then the potential
$W$ in (4.14 ) is given by (see (2.31),(4.13) ; we shall assume that $d_1$
is positive)
$$ W = \BW / 2N \ \ , \ \ \
\BW= d_1 \e{ \vp + N\l}    \ ,   $$
and our system takes the form ( the terms proportional to $c$
can be neglected in the large $\l$ limit )
$$ -N\dls +\dvps = \BW \ , \ \ \ddl - \dvp \dl = -\half \BW \ , \ \ \ddvp - N
\dls = \half \BW \ . \eq{B.1}  $$
Combining these equations we get
$$ \ddl  - \ddvp = \dvp ( \dl - \dvp ) \ , \ \ \ \ \dl = \dvp + A \e{\vp}  \ ,
\eq{B.2} $$
$$ \  \ddvp = \half ( N \dls + \dvps ) \ \ , \eq{B.3} $$
To have an expansion with decreasing dilaton , i.e. $ {\dot \l } > 0 \ ,
\ {\dot \vp } < 0 $
we should take $A < 0$ . Substituting the expression for $\dl$ in (B.2)
into (B.3) we get the  equation for $\vp$ which is easy to solve by
introducing the new time variable $\tau$
$$ d\tau =  dt\  \e{\vp} \ \  , \ \ \vp' = {d\vp \over d \tau } = \e{-\vp}
\dvp \  \ , \ \ \l' = \vp' + A \  \ ,  \eq{B.4} $$
so that
$$ \vp'' = \half [ N ( \vp ' + A )^2 - \vp'^2 ] \ \ . \eq{B.5} $$
Let us first consider the case of $N > 1$ .
 Integrating (B.5) we find
$$\vert {\vp' + u \over \vp' + v} \vert = \e{-\sn A \tau } \ \ ,
\eq{B.6} $$
$$ u={\sn A \over \sn -1 } \ \ , \ \ v = { \sn A \over \sn +1 } \ \
. \eq{B.7} $$
As a result  ( we assume that $-u < \vp' < - v $ )
$$ \vp' = - { u + v z \over 1 + z } \ \ , \ \ z \equiv \e{- \sn A
\tau } \ \ ,$$
$$ \vp = \vp_0 + {1\over \sn -1 } {\ln }\ z - {2\over N-1 } {\ln }\
(1+z) \ \ ,  \eq{B.8} $$
 $$ \l = \l_0 + {1\over \sn (\sn -1 )} {\ln} \ z - {2\over N-1}{ \ln }\
(1+z) \ \ .  \eq{B.9} $$
The first equation in (B.2)
$$ \vp'^2 - N \l'^2 = d_1 \e{-\vp + N \l} \ \ ,
\eq{B.10} $$
 is then satisfied if
$$ A^2 = {1\over 4N } (N-1 ) d_1 \ \ . \eq{B.11} $$
 The relation between the original time $t$ and $z$ is
$$ dt = d \tau \e{-\vp} = - ( N A^2 )^{-1/2} dz z^{a} (1+z)^b\ , \ \
a= -{\sn \over \sn -1 } \ , \ \
b={2\over N-1} \ \ , \eq{B.12} $$
which can be approximated by $t \sim z^{-h} = {\exp (\sn Ah\tau ) }
\ , \ \  0< h < 1 $ , i.e. $z$ decreasing to zero corresponds to $t$
growing to infinity.

As it is clear from (B.9) , $\ \l (z)$ first
grows to its maximal value at $z_* = {\sn+1 \over \sn -1}
$ and then decreases. The function $\vp (z)$ is monotonically
decreasing with $z \ra 0$ .  We conclude that a maximal radius of
 expansion is reached in a finite time  ( $z_*$ corresponds to a
turning point of the trajectory ). Once the contraction starts we
eventually reach the region of
 small $\l$  where  other terms in $W$ (4.14)
 (which  are present because of the contributions the winding modes)
  cannot be ignored .

If $N=1$  the integration of (B.5) gives
$$\vp=\vp_0 - \half A \tau + A^{-1}{ \rm e}^{A\tau} \ , \ \ \l= \l_0 + \half A
 \tau
+ A^{-1} {\rm e}^{A \tau} \ . \eq{B.13}  $$
A different way of solving (B.1) in the $N=1$ case was discussed
in Sect.4 (see (4.23)--(4.25)).


\vfill\eject

\centerline{\bf References}

{\settabs 16\columns
\+ 1. & K. Kikkawa and M. Yamasaki,
      Phys. Lett. B149 (1984) 357 \cr
\+    & N. Sakai and I. Senda,  Progr. Theor. Phys. 75(1986)692\cr
\+    & V. Nair, A. Shapere, A. Strominger and F. Wilczek,
   Nucl. Phys. B287(1987)402\cr
\+ 2. & R. Brandenberger and C. Vafa, \NP B316(1988) 391 \cr
\+ 3. & H. Nishimura and M. Tabuse, \MPL A2(1987)299 \cr
\+ 4. & J. Kripfganz and H. Perlt, Class. Quant. Grav. 5(1988)453 \cr
\+ 5. & T.H. Buscher, \PL B194(1987)59 ; \PL B201(1988)466 \cr
\+    & G. Horowitz and A.R. Steif, \PL B250(1990)49 \cr
\+ 6. & P. Ginsparg and C. Vafa,  Nucl. Phys. B289(1987)414\cr
\+    & A. Giveon, N. Malkin and E. Rabinovici, Phys. Lett. B220(1989)551\cr
\+    & E. Alvarez and M. Osorio, Phys. Rev. D40(1989)1150\cr
\+ 7. & T. Banks, M. Dine, H. Dijkstra and W. Fischler, \PL B212(1988) 45
\cr
\+ 8. & E. Smith and J. Polchinski, \PL B263(1991)59  \cr
\+ 9. & A.A. Tseytlin, \MPL A6(1991)1721 \cr
\+ 10.& G. Veneziano, preprint CERN-TH-6077/91 \cr
\+ 11.& A.A. Tseytlin, ``Space--time duality, dilaton and string
cosmology'', \cr
\+    & Proc. of the First International A.D. Sakharov
Conference on Physics, Moscow \cr
\+    &  27- 30 May 1991, ed.  L.V. Keldysh
et al., Nova Science Publ., Commack,
N.Y. , 1991 \cr
\+ 12.& A.A. Tseytlin and C. Vafa, Harvard preprint HUTP-91/A049 \cr
\+ 13.&J.V. Jose, L.P. Kadanoff, S. Kirkpatrick and D. Nelson,
 \PR B16(1977)1217 \cr
\+    & P. Wiegman, J. Phys. C11(1978)1583 \cr
\+    & D.J. Amit, Y.Y. Goldschmidt and G. Grinstein, J.Phys. A13(1980)585
\cr
\+    & D. Boyanovski and R. Holman, Nucl. Phys. B332(1990)641 \cr
\+ 14.& R. Myers, \PL B199(1987)371 \cr
\+    & I. Antoniadis, C. Bachas, J. Ellis and D. Nanopoulos,
\PL B211(1988)393;\cr
\+    & \NP B328(1989)115 \cr
\+ 15.& M. Mueller, \NP B337(1990)37 \cr
\+ 16.& N. Sanchez and G. Veneziano, \NP B333(1990)253 \cr
\+ 17.& B.A. Campbell, A. Linde and K.A. Olive, \NP B355(1991)146  \cr
\+ 18.& A. Cooper, L. Susskind and L. Thorlacius, \NP B363(1991)132 \cr
\+ 19.& J. Scherk and J.H. Schwarz, Nucl. Phys. B81(1974)118 \cr
\+    & E.S. Fradkin and A.A. Tseytlin, \NP B261(1985)1 \cr
\+    & C.G. Callan, D. Friedan, E. Martinec and M.J. Perry, \NP
B262(1985)593 \cr
\+ 20.& A.A. Tseytlin, \NP B350(1991)395 \cr
\+ 21.& J. Polchinski, Commun. Math. Phys. 104(1986)37 \cr
\+ 22.& M. Hellmund and J. Kripfganz, \PL B241(1990)211 \cr
\+ 23.& D.J. Gross and I.R. Klebanov, \NP B344(1990)475 \cr
\+    & M. Bershadsky and I.R. Klebanov, \PRL 65(1990)3088 \cr
\+    & N. Sakai and Y. Tanii, Int. J. Mod. Phys. A6(1991)2743   \cr

\vfill
\bye